\documentclass{iau_FM}

\usepackage{graphicx}

\begin{document}
\title[FM 8.~~Optimizing Faraday Grids] 
{Optimizing Faraday Background Grids}

\author[Lawrence Rudnick]   
{Lawrence Rudnick$^1$}
\affiliation{Minnesota Institute for Astrophysics, University of Minnesota, 116 Church St. SE, Minneapolis, MN 55455 USA  email: {\tt larry@umn.edu}}
\pubyear{2019}
\setcounter{page}{1}
\jname{XXX} 
\editors{XXX, ed.}
\maketitle
\begin{abstract}
Magnetic field strengths in objects ranging from  HII regions to cosmological large scale structure can be estimated using dense grids of Rotation Measures (RMs) from polarized background radio structures.  Upcoming surveys on the SKA and its precursors will dramatically increase the number $N$ of background sources.  However, detectable magnetic field strengths will scale only as $t^{-0.15}$, for an integration time $t$ on a fixed area of sky, so the analysis techniques need to be optimized.  A key factor is the difference in the dispersion of intrinsic RMs for different populations, which must be carefully accounted for to achieve the scientifically needed accuracies.
\end{abstract}

\firstsection 
\section{Introduction}
The next generation of radio surveys, including the SKA and its precursors, will dramatically increase the number of detected distant polarized sources. When these sources are viewed through a foreground magnetized plasma, whether it is from an HII region, a galaxy, or cluster of galaxies, or even large scale structure, Faraday effects in that plasma offer a way to explore its magnetic field strength and structure (\cite{jh}). \cite{aka} describes the many contributions to the observed rotation measures (RMs), from which the effects of the intervenor must be isolated.  The accuracy with which this can be done, as detailed below, depends on the number of background sources and the scatter in their RMs.

\begin{figure}[h]
\begin{center}
 \includegraphics[width=2.in]{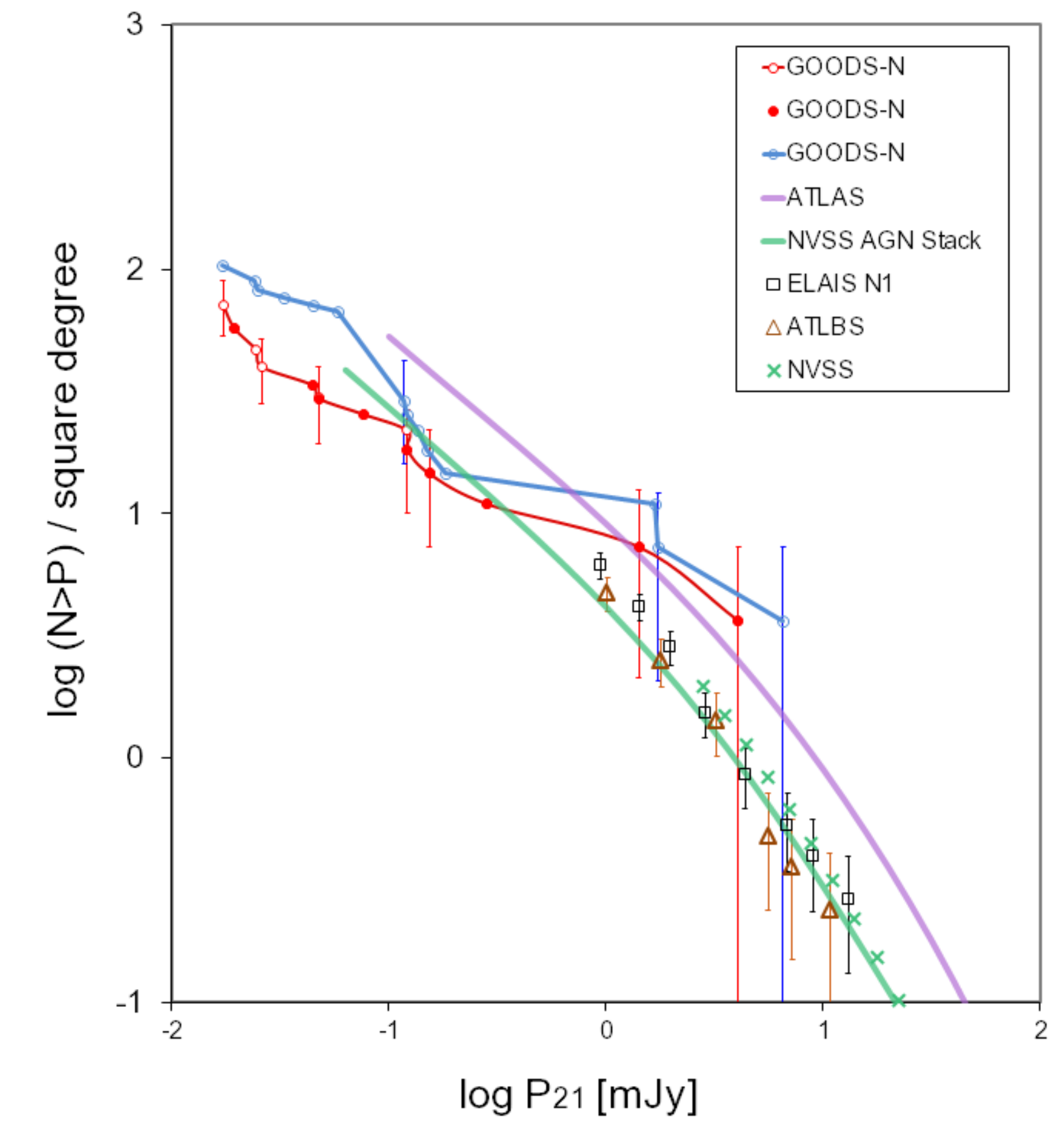}
 \caption{Cumulative polarization counts.  Adapted from \cite{goods}.}
   \label{fig1}
\end{center}
\end{figure}

Increasing the number of polarized sources is expensive, as seen in Figure \ref{fig1}, since the cumulative number increases only as $P_{det}^{-0.6}$ where $P_{det}$ is the detection limit.  This requires that we optimize our techniques for using the limited number of sources that we have; one aspect of that optimization is discussed here.
\section{Optimizing the figure of merit }
In background grid experiments we measure the difference in the variance of RMs seen through a foreground screen to those of a control sample.  The sensitivity thus depends on the intrinsic RM variance of individual background sources (the smaller the better), and the number that are available behind the screen (the more the better).   

Any such experiment will have multiple populations of background sources.  If sources of type $j$ have individual values of $RM_{j,i}$, a sky density of $n_j$ /steradian, and our foreground object of interest covers $\Omega$ steradians, then we have $N_j = n_j \Omega$ sources.   The population will have a variance in RM, $\sigma_j^2$, and a corresponding uncertainty, of 
$$\sigma_j^2 = \langle RM_{j,i}^2\rangle. - \langle RM_{j,i}\rangle^2 ~ ~ \textrm{and}~ ~  \delta_{j}\approx \sqrt{\frac{2}{N_j}}~\sigma_j^2 ~ ~   \frac{rad^2}{m^4}$$
for a Gaussian distribution of $N_j$ sources (\cite{stats}).
$\delta_j$ is the error in $\sigma_j^2$, the figure of merit that we would like to minimize. \textbf{It is important to note, and not generally recognized, that this leads to a minimum magnetic field which scales only as $P_{det}^{0.3}$ or an integration time of $t^{-0.15}$ for observations of a single area on the sky. 
}.  This general behavior applies whether one is using \textit{F-tests} or Bayesian statistics, or Monte-Carlo type modeling, etc.

As an example, consider two populations, $a$ and $b$, with different values of $\sigma_a^2$ and $\sigma_b^2$, and  the same sample sizes, $N$.   The default procedure is just to average them.  Then  $$\sigma_{tot,unwtd}^2= 0.5*(\sigma_a^2+\sigma_b^2)~ ~  \textrm{and} ~ ~   \delta_{tot,unwtd}= 0.5*(\sqrt{\delta_a^2 + \delta_b^2}) $$.   However, if we were to weight by the inverse variances of the two populations, then $$\sigma_{tot,wtd}^2= (\frac{\sigma_a^2}{\delta_a^2}+\frac{\sigma_b^2}{\delta_b^2})/(\frac{1}{\delta_a^2}+\frac{1}{\delta_b^2})~ \textrm{and}~  \delta_{tot,wtd} = \frac{1}{\sqrt{(\frac{1}{\delta_a^2}+\frac{1}{\delta_b^2})}}$$.

To illustrate the difference, assume that $\delta_b = 10 \delta_a$, and that the numbers of  sources are the same (N) in the two populations, so  $\sigma_0 \equiv \sigma_b = \sqrt{10}~\sigma_a$.  As we will see below, such differences in RM variations between different populations are common.  Now, 
$$\delta_{tot,unwtd}=2.3 \sigma_0^2 /\sqrt{N} ~ ~ \textrm{and} ~ ~ \delta_{tot,wtd}=0.48 \sigma_0^2 /\sqrt{N}$$

\begin{figure}[]
\begin{center}
 \includegraphics[width=2.5in]{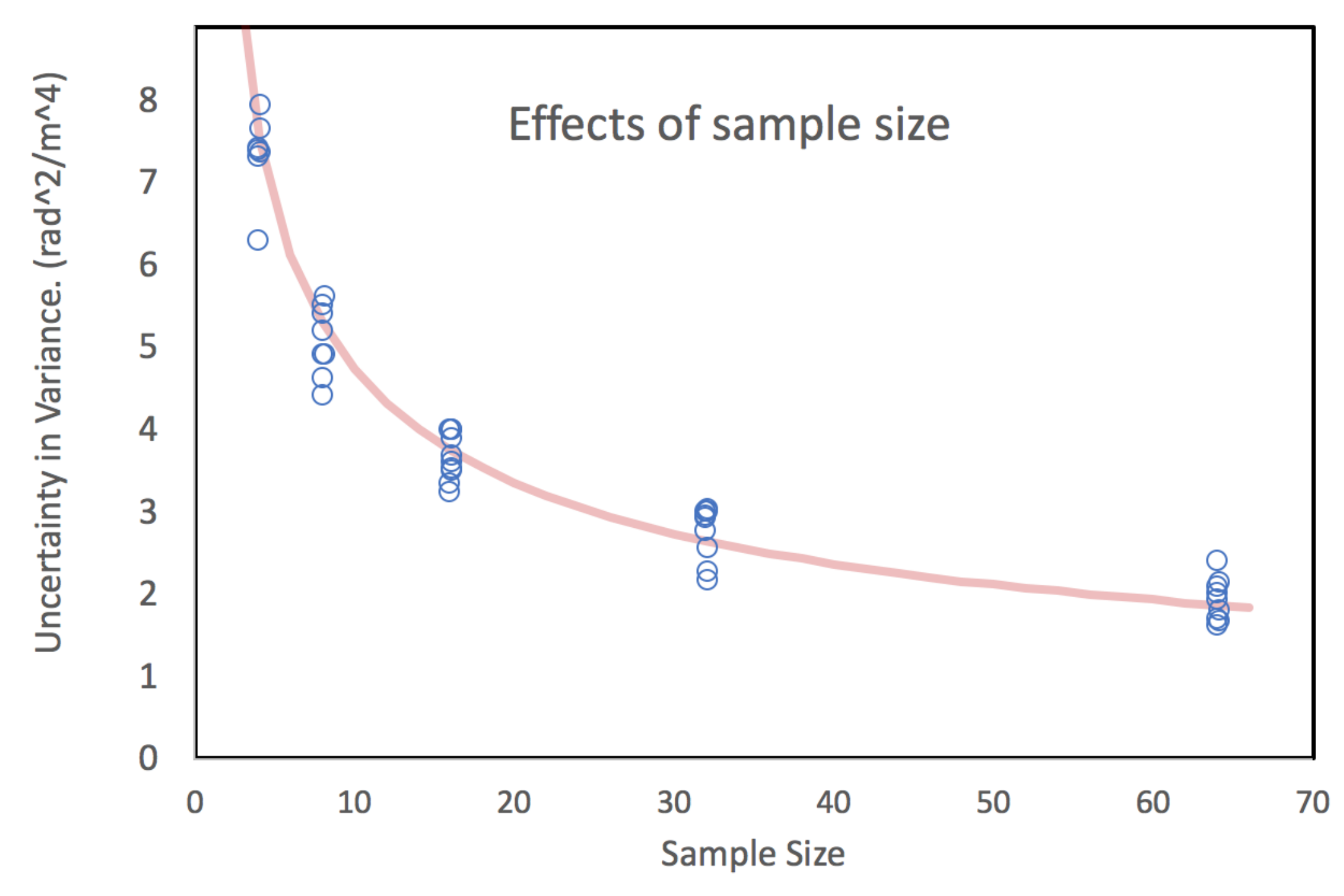} 
 \includegraphics[width=2.5in]{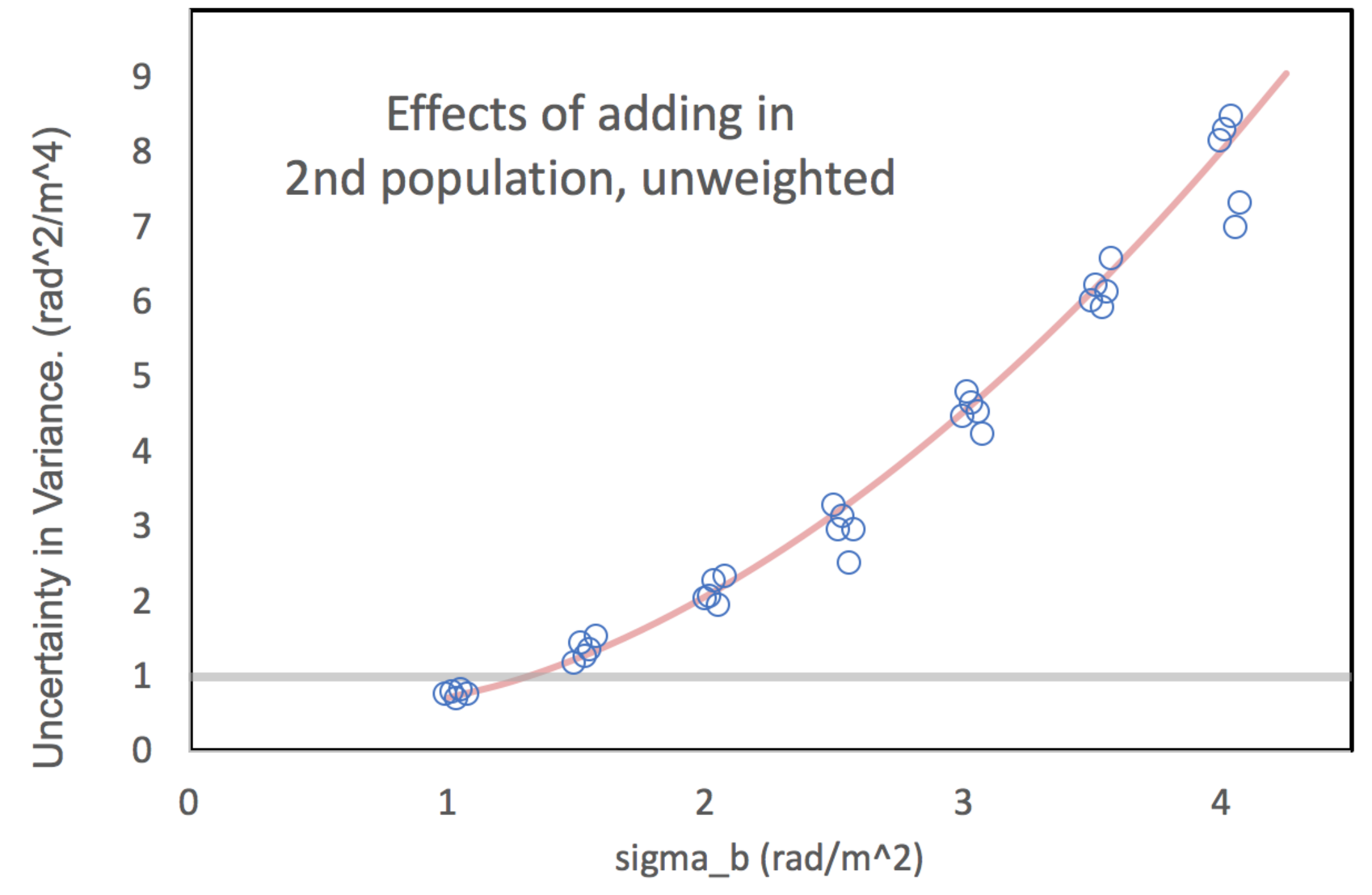} 
 \caption{Uncertainty in \textbf{variance} for background experiments.  Left: Varying the number of sources in each population.  Right: \textbf{Unweighted} analysis as a function of rms scatter of 2nd population (1st population = 1).}
   \label{variance}
\end{center}
\end{figure}

Weighting improves the accuracy of the variance by a factor of over 4 in this case.  In Figure \ref{variance} we show the results of this type of experiment, varying $N$ and varying the ratio of $\frac{\sigma_b}{\sigma_a}$.   The $\frac{1}{\sqrt{N}}$ behavior is seen on the left. On the right, we see that when the two population variances are equal, one gets an \textit{improvement} by a factor of $\sqrt{2}$, as expected.  But as $\sigma_b$ rises, even though we are adding more sources, the uncertainty in the unweighted variance goes up. Weighting is critical! 
This calculation should be redone both for verification and to put in realistic models of the distributions of RMs in different populations, since they are quite unlikely to have Gaussian distributions.

\section{Populations and variances}
Which source properties will lead to different RM variances?  \textit{All of them!}

\underline{Galactic location} (and especially) latitude is the first major factor.  Rotation measures and the scatter among them increase strongly at low galactic latitudes and in the direction of some local Faraday structures.  For studies of individual objects, this means that control samples must be done in close proximity to the foreground screen of interest.   For studies of large samples, the different variances due to screens and background sources at different galactic locations must be properly weighted.

The nature of the \underline{optical host} leads to very different polarization properties.  \cite{osull} found large differences in the fractional polarizations of radiative-mode and jet-mode AGN, which could be related to Faraday differences (see below).

The \underline{spectral index} of the background radio source influences its polarization properties.  \cite{farnes} shows that the ``(fractional) polarization spectral index", differs for flat and steep spectrum sources.  Negative slopes indicate depolarization, i.e., fractional polarization decreasing with increasing wavelength.  Often, depolarization will be accompanied by a variation in RM as a function of wavelength.  This variation will increase the intrinsic variance in RM within the population.  When the polarization spectral index is positive, i.e., the source "re-polarizes" at longer wavelengths; then there will always be a variation of RM with wavelength, and thus an increased RM variance.
\vspace{-0.1cm}
\subsection{Fractional Polarization}
The integrated fractional polarization of a radio source primarily depends on two factors, \textit{i)} variations of the magnetic field direction within the synchrotron emitting source, and \textit{ii)} variations in the Faraday depth either within or across the observed structure of the source.  \textit{i)} does not lead to any particular RM behavior, while \textit{ii)} will create strong correlations between fractional polarization and depolarization. Initial indications are that both of these can be important.  So fractional polarization is an imperfect indicator of RM variations --  a very high fractional polarization implies small RM variations, while a low fractional polarization could be due to either effect.
\begin{figure}[h]
\begin{center}
 \includegraphics[width=2.4in]{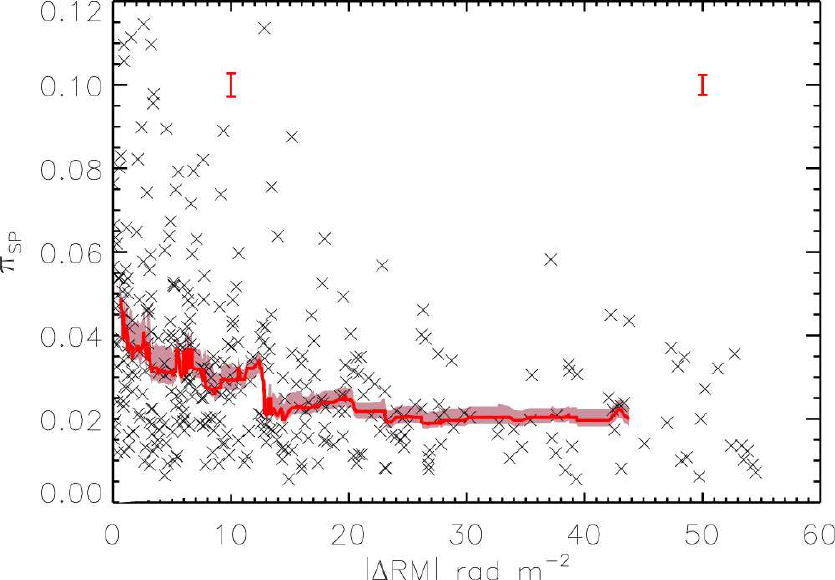}
 \caption{Relation between RM variations and fractional polarization,  \cite{lamee}.}
   \label{pivRM}
\end{center}
\end{figure}
Figure \ref{pivRM} shows the relationship between RM variations and fractional polarizations of sources selected from the S-PASS survey (\cite{lamee}).  $\Delta RM$ is the difference between RMs measured between a)  1365~MHz and 1435~MHz from the NVSS survey and b) between 1400~MHz and 2300~MHz from the NVSS and S-PASS surveys.  As long as $\Delta RM \neq 0$, this will result in an extra contribution to the RM variations in a population.  The correlation between fractional polarization and $\Delta RM$ shows that low polarization sources are much more likely to have large RM population variances.

Fractional polarization itself is a strong function of both the angular and physical size of the source (work in progress) as well as the total intensity.  While this doesn't guarantee that RM variances will depend on these quantities, correcting for any dependencies on size and intensity are likely to be critical.

\subsection{Morphology}
\begin{figure}[h]
\begin{center}
 \includegraphics[width=2.5in]{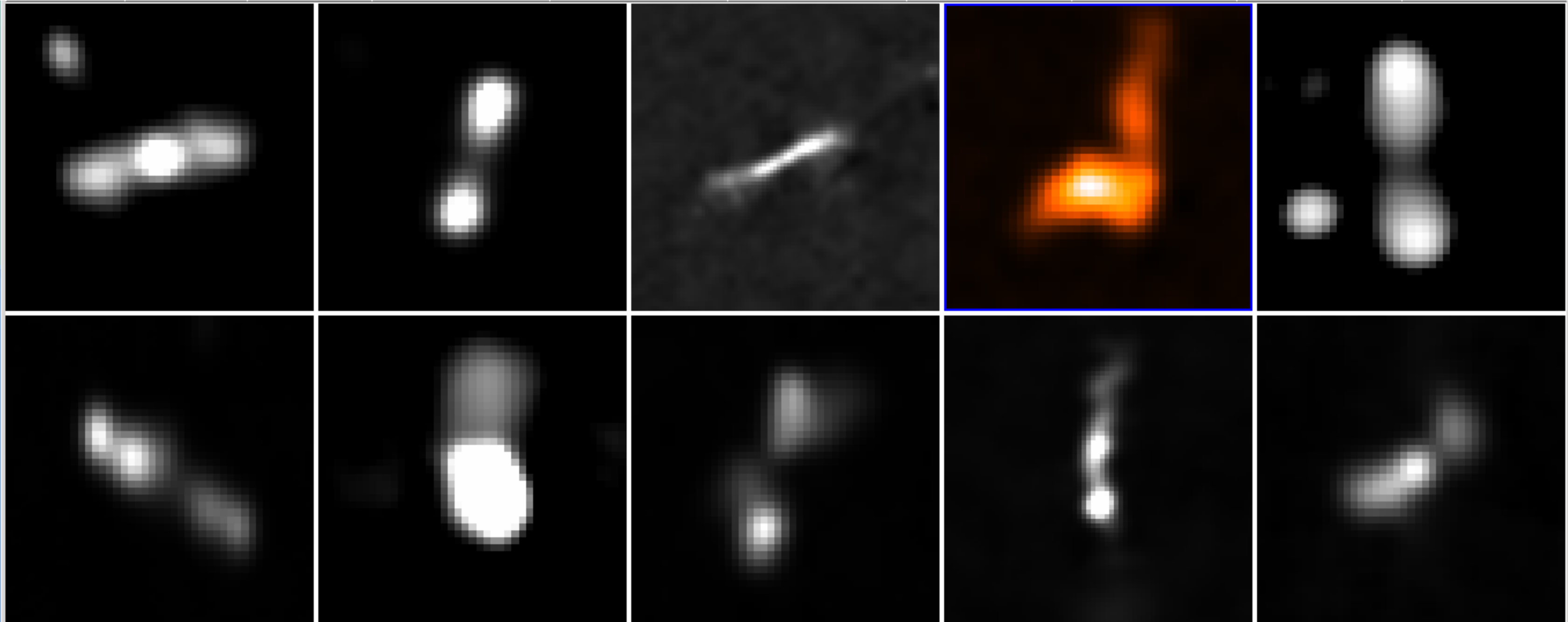}
  \includegraphics[width=2.5in]{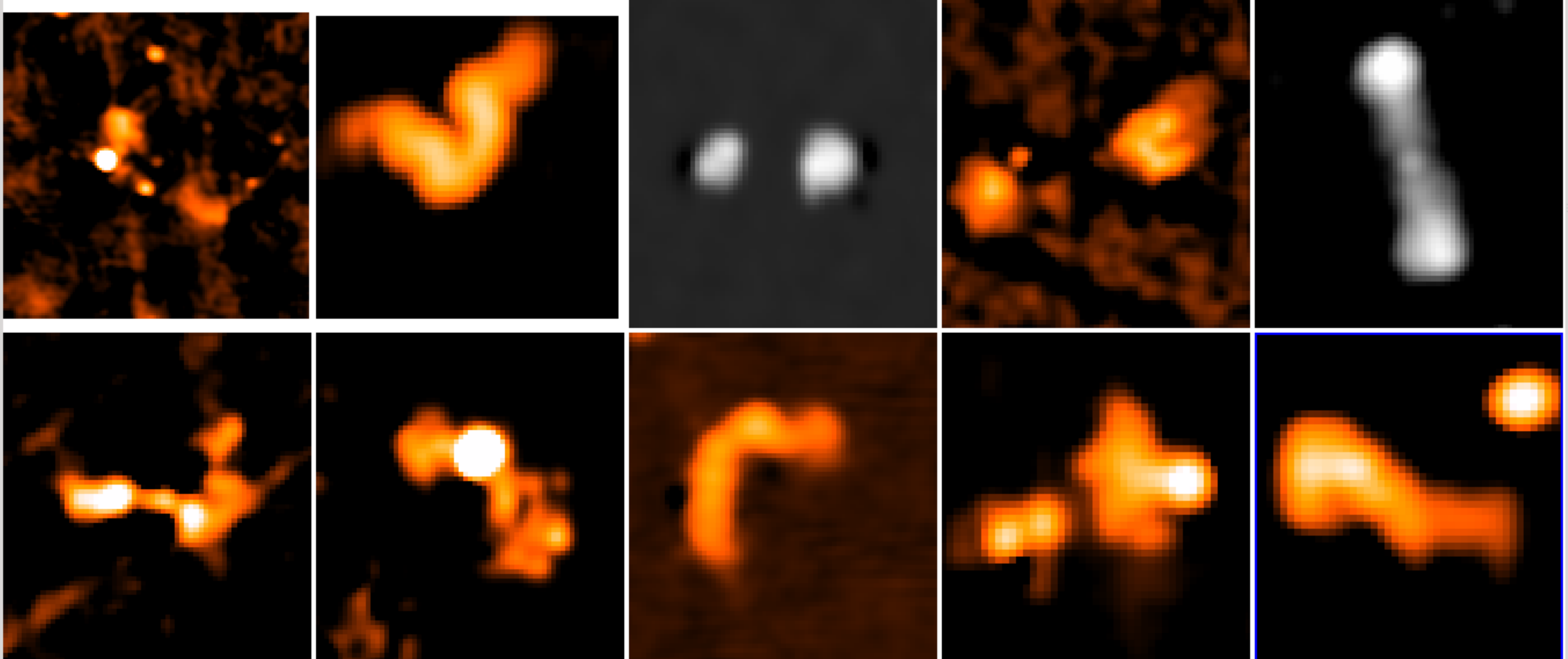}
 \caption{The NVSS structure of low (left) and high (right) $\sigma_{RM}$ sources.}
   \label{morph}
\end{center}
\end{figure}

There is also a strong relation between source morphology and the strength of RM variations across the source, although this has not been systematically investigated to our knowledge.  An initial look at this issue has been made by M. Wieber (Minnesota), based on samples of large angular size sources assembled by H. Andernach (Guanajuato) and S. O'Sullivan (Hamburg).   These sources are resolved in NVSS, and polarization maps in the two NVSS bands provided by J. Stil (Calgary) were used to characterize $\sigma_{RM}$, the variation in RM across the face of the source.  Representative groups of the lowest $\sigma_{RM}$ sources and the highest ones are shown in Figure \ref{morph}.  Color coding has been used to guide the eye, but it is clear that sources with larger values of $\sigma_{RM}$ are more likely to show structures with bends and other distortions.   This relationship is not surprising, given that the distorted sources are interacting with a substantial external thermal medium, with the higher densities (and perhaps fields) leading to higher RMs.
\vspace{ -0.1in}
\section{Conclusions}
For studies of magnetic fields in foreground screens, using populations of polarized background sources, it is critical to measure the intrinsic variations in the RM distributions of each background population, and properly weight them in the analysis. \textit{This work is supported, in part, by U.S. NSF grant AST17-14205 to the University of Minnesota.}


\vspace{-0.1in}
\begin{thebibliography}{}
\bibitem[Akahori, Gaensler \& Ryu (2014)]{aka}
Akahori, T., Gaensler, B. \& Ryu, D., 2014, \textit{ApJ} 790, 123
\bibitem[Farnes et al. (2014)]{farnes}
Farnes, J., Gaensler, B. \& Carretti, E., 2014, \textit{ApJSS} 212, 15
\bibitem[Johnston-Hollit et al. (2014)]{jh}
Johnston-Hollit, M. et al. 2014, \textit{POS-AASKA14}, https://pos.sissa.it/215/092/pdf
\bibitem[Lamee et al. (2016)]{lamee}
Lamee, Mehdi; Rudnick, Lawrence; Farnes, Jamie S.; Carretti, Ettore; Gaensler, B. M.; Haverkorn, Marijke; Poppi, Sergio 2016, \textit{ApJ} 829, 18
\bibitem[Lehman \& Casella (1998)]{stats}
Lehman, E. \& Casella, G., \textit{Theory of point estimation}, Springer-Verlag, New York
\bibitem[O'Sullivan et al. (2014)]{osull}
O'Sullivan, S., et al.,   2017, \textit{MNRAS} 469, 4034
\bibitem[Rudnick \& Owen (2014)]{goods}
Rudnick, L. \& Owen, F. N.  2014, \textit{ApJ} 785, 45
\end{thebibliography}
\end{document}